# A Progressive Mesh Method for Physical Simulations Using Lattice Boltzmann Method on Single-Node multi-GPU Architectures

Julien Duchateau[1], François Rousselle[1], Nicolas Maquignon[1], Gilles Roussel[1], Christophe Renaud[1]

[1]Laboratoire d'Informatique, Signal, Image de la Côte d'Opale
Université du Littoral Côte d'Opale, Calais, France

*ABSTRACT*

*In this paper, a new progressive mesh algorithm is introduced in order to perform fast physical simulations by the use of a lattice Boltzmann method (LBM) on a single-node multi-GPU architecture. This algorithm is able to mesh automatically the simulation domain according to the propagation of fluids. This method can also be useful in order to perform several types of physical simulations. In this paper, we associate this algorithm with a multiphase and multicomponent lattice Boltzmann model (MPMC–LBM) because it is able to perform various types of simulations on complex geometries. The use of this algorithm combined with the massive parallelism of GPUs[5] allows to obtain very good performance in comparison with the staticmesh method used in literature. Several simulations are shown in order to evaluate the algorithm.*

## 1. INTRODUCTION

The lattice Boltzmann method (LBM) is a computational fluid dynamics (CFD) method. It is a relatively recent technique which is able to approximate Navier-Stokes equations by a collision-propagation scheme [1]. Lattice Boltzmann method however differs from standard approaches as finite element method (FEM) or finite volume method (FVM) by its mesoscopic approach. It is an interesting alternative which is able to simulate complex phenomena on complex geometries. Its high parallelization makes also this method attractive in order to perform simulations on parallel hardware. Moreover, the emergence of high-performance computing (HPC) architectures using GPUs [5] is also a great interest for many researchers.

Parallelization is indeed an important asset of lattice Boltzmann method. However, perform simulations on large complex geometries can be very costly in computational resources. This paper introduces a new progressive mesh algorithm in order to perform physical simulations on complex geometries by the use of a multiphase and multicomponent lattice Boltzmann method. The algorithm is able to automatically mesh the simulation domain according to the propagation of fluids. Moreover, the integration of this algorithm on single-node multi-GPU architecture is also an important matter which is studied in this paper. This method is an interesting alternative which has never been exploited at the best of our knowledge.

Section 2 first describes the multiphase and multicomponent lattice Boltzmann method. It is able to simulate the behavior of fluids with several physical states (phase) and it is also able to model several fluids (component) interacting with each other. Section 3 presents then several recent works involving lattice Boltzmann method on GPUs. Section 4 mostly concerns the main contribution of this paper: the inclusion of a progressive mesh method in the simulation code. The principles of the method and the definition of an adapted criterion are firstly introduced. The integration on a single-node multi-GPU architecture is then described. An analysis concerning performance is also studied in section 5. The conclusion and future works are finally presented in the last section.

## 2. THE LATTICE BOLTZMANN METHOD

### 2.1. The Single relaxation time Bhatnagar-Gross-Krook (SRT-BGK) Boltzmann equation

The lattice Boltzmann method is based on three main discretizations: space, time and velocities. Velocity space is reduced to a finite number of well-defined vectors. Figures 1(a) and 1(b) illustrate this discrete scheme for D2Q9 and D3Q19 model.

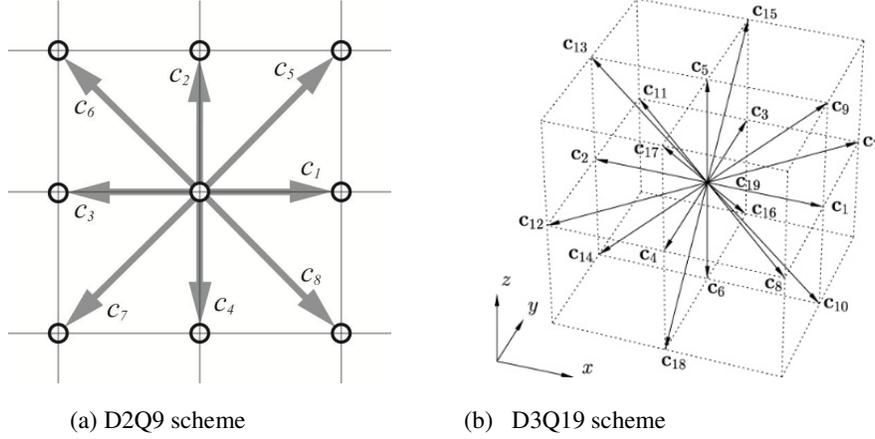

(a) D2Q9 scheme       (b) D3Q19 scheme

Figure 1: Example of Lattice Boltzmann schemes

The simulation grid is therefore discretized as a Cartesian grid and calculation steps are achieved on this entire grid. The discrete Boltzmann equation[1] with a single relaxation time Bhatnagar-Gross-Krook (SRT-BGK) collision term is defined by the following equation:

$$f_i(x + e_i, t + \Delta_t) - f_i(x,t) = \frac{1}{\tau}\left(f_i(x,t) - f_i^{eq}(x,t)\right) \quad (1)$$

$$f_i^{eq}(x,t) = \omega_i \rho(x,t)\left(1 + \frac{e_i u}{c_s^2} + \frac{(e_i u)^2}{2c_s^4} - \frac{u^2}{2c_s^2}\right) \quad (2)$$

$$c_s^2 = \frac{1}{3}\left(\frac{\Delta_x}{\Delta_t}\right)^2 \quad (3)$$

The function $f_i(x,t)$ corresponds to the discrete density distribution function along velocity vector $e_i$ at a position $x$ and a time $t$. The parameter $\tau$ corresponds to the relaxation time of the simulation. The value $\rho$ is the fluid density and $u$ corresponds to the fluid velocity. $\Delta_x$ and $\Delta_t$ are the spatial and temporal steps of the simulation respectively. Parameters $w_i$ are weighting values defined according to the lattice Boltzmann scheme and can be found in [1]. Macroscopic quantities as density $\rho$ and velocity $u$ are finally computed as follows:

$$\rho(x,t) = \sum_i f_i(x,t) \tag{4}$$

$$\rho(x,t)u(x,t) = \sum_i f_i(x,t)e_i \tag{5}$$

## 2.2. Multiphase and Multi Component Lattice Boltzmann Model

Multiphase and multicomponent models (MPMC) allow performing complex simulations involving several physical components. In this section, a MPMC-LBM model based on the work achieved by Bao & Schaeffer [4] is presented. It includes several interaction forces based on pseudo-potential. It is calculated as follows:

$$\psi_\alpha = \sqrt{\frac{2(p_\alpha - c_s^2 \rho_\alpha)}{c_s^2 g_{\alpha\alpha}}} \tag{6}$$

The term $p_\alpha$ is the pressure term. It is calculated by the use of an equation of state as the Peng-Robinson equation:

$$p_\alpha = \frac{\rho_\alpha R_\alpha T_\alpha}{1 - b_\alpha \rho_\alpha} - \frac{a_\alpha \theta(T_\alpha)\rho_\alpha^2}{1 + 2b_\alpha - b^2\rho^2} \tag{7}$$

Internal forces are then computed. The internal fluid interaction force is expressed as follows [2] [3]:

$$F_{\alpha\alpha}(x) = -\beta \frac{g_\alpha}{2} c_s^2 \psi_\alpha(x) \sum_{x'} w_i \psi_\alpha(x')(x'-x) - \frac{1-\beta}{2} \frac{g_\alpha}{2} c_s^2 \psi_\alpha(x) w_i \psi_\alpha^2(x')(x'-x) \tag{8}$$

The value $\beta$ is a weighting term generally fixed to 1.16 according to [2] [3]. The inter-component force is also introduced as follows [4]:

$$F_{\alpha\alpha'}(x) = -\frac{g_{\alpha\alpha'}}{2} c_s^2 \psi_\alpha(x) \sum_{x'} w_i \psi_\alpha(x')(x'-x) \tag{9}$$

Additional forces can be added into the simulation code as the gravity force, or a fluid-structure interaction [3]. The incorporation of the force term is then achieved by a modified collision operator expressed as follows:

$$f_{\alpha,i}(x + e_i, t + \Delta_t) - f_{\alpha,i}(x,t) = \frac{1}{\tau}\left(f_{\alpha,i}(x,t) - f_{\alpha,i}^{eq}(x,t)\right) + \Delta f_{\alpha,i} \tag{10}$$

$$\Delta f_{\alpha,i} = f_{\alpha,i}^{eq}(\rho_\alpha, u_\alpha + \Delta u_\alpha) - f_{\alpha,i}^{eq}(\rho_\alpha, u_\alpha) \tag{11}$$

$$\Delta u_\alpha = \frac{F_\alpha \Delta_t}{\rho_\alpha} \tag{12}$$

Macroscopic quantities for each component are finally computed by the use of equations (4) and (5).

## 3. LATTICE BOLTZMANN METHODS AND GPUS

The mass parallelism of GPUs has been quickly exploited in order to perform fast simulations [7] [8] using lattice Boltzmann method. Recent works have shown that GPUs are also used with multiphase and multicomponent models [16] [14]. The main aspects of GPU optimizations are decomposed into several categories [10] [9] as thread level parallelism, GPU memory access, overlap of memory transfers with computations …. Data coalescence is needed in order to optimize global memory bandwidth. This implies several conditions as described in [9]. Concerning LBM, an adapted data structure such as the Structure of Array (SoA) has been well studied and has proven to be efficient on GPU [7].

Several access patterns are also described in the literature. The first one, named A-B access pattern, consists of using two calculation grids in GPU global memory in order to manage the temporal and spatial dependency of the data (Equation (10)). Simulation steps alternate between reading distribution functions from A and writing them to B, and reading from B and writing to A reciprocally. This pattern is commonly used and offers very good performance [10] [11] [9] on a single GPU. Several techniques are however presented in literature in order to reduce significantly the computational memory cost without loss of information such as grids compression [6], Swap algorithm [6] or A-A pattern technique [12]. In this paper, the A-A pattern technique is used in order to save memory due to spatial and temporal data dependency.

Recent works involving implementation of lattice Boltzmann method on a single-node composed of several GPUs are also available. A first solution, proposed in [13] [17], consists in dividing the entire simulation domain into subdomains according to the number of GPUs and performing LBM kernels on each sub-domain in parallel. CPU threads are used to handle each CUDA context. Communications between sub-domains are performed using zero-copy memory transfers. Zero-copy feature allows to perform efficient communications by a mapping between CPU and GPU pointers. Data must however be read and written only once in order to obtain good performance.

Some approaches have finally been proposed recently to perform simulations on several nodes constituted of multiple GPUs by the use of MPI in combination with CUDA [19][18][21] [15]. In our case, we only dispose of one computing node with multiple GPUs thus we don't focus on these architectures in this paper.

## 4. A PROGRESSIVE MESH ALGORITHM FOR LATTICE BOLTZMANN METHODS ON SINGLE-NODE MULTI-GPU ARCHITECTURES

### 4.1. Motivation

Works described in the previous section consider that the entire simulation domain is meshed and divided into subdomains according to the number of GPUs, as shown on Figure 2. All subdomains are therefore calculated in parallel.

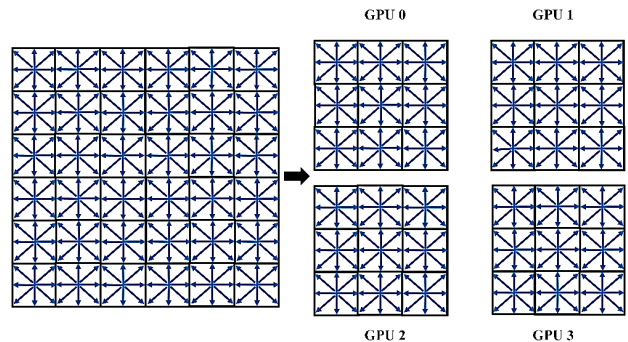

Figure 2: Division of the simulation domain: the entire domain is decomposed into subdomains according to the number of GPUs.

In this paper, a new approach is considered. For most simulations, the entire domain generally does not requires to be fully meshed at the beginning of the simulation. We propose therefore a new progressive mesh method in order to dynamically create the mesh according to the propagation of the simulated fluid. The idea consists in defining a first subdomain at the beginning of the simulation (Figure 3(a)). Several subdomains can then be created following the propagation of the fluid as can be seen of Figure 3(b). This method finally adapts automatically to the simulation geometry (Figure 3(c)). This method is therefore applicable for any geometry and simulations. It is also a real advantage for an application on industrial structures mostly composed of pipes or channels. It can indeed save a lot of memory and calculations according to the geometry used for the simulation.

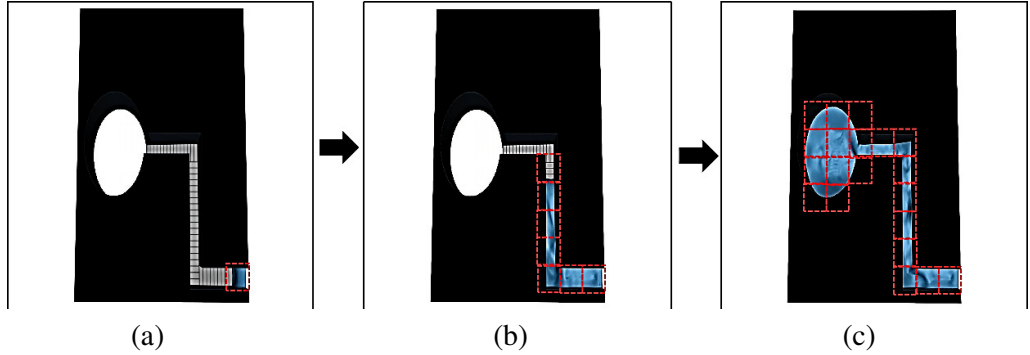

(a)  (b)  (c)

Figure 3: Example of a 3D simulation using the progressive mesh algorithm: (a) a first subdomain is created at the beginning of the simulation, (b) several subdomains are created following the propagation of fluid, (c) all subdomains are created and completely adapt to the simulation geometry.

The progressive mesh algorithm firstly needs the introduction of an adapted criterion in order to create a new subdomain to the simulation. This new subdomain needs then to be connected to existing subdomains. Calculations on single-node multi-GPU architecture are finally an important optimization factor.

### 4.2. Definition of a Criterion for the Progressive Mesh

The definition of a criterion is an important aspect in order to efficiently create new subdomains for the simulation. This criterion needs to represent efficiently the propagation of fluid. The fluid velocity seems like a good choice in order to define an efficient criterion. The difference of the fluid velocity between two iterations is considered in order to observe efficiently the fluid dispersion. Our criterion is therefore defined as follows for the component $\alpha$:

$$\|C_\alpha(x)\|_2 = \|u_\alpha(x, t + \Delta_t) - u_\alpha(x, t)\|_2 \quad (13)$$

The symbol $\|.\|_2$ stands for the Euclidean norm in this paper. This criterion needs to be calculated for all active subdomains on the boundaries. If the criterion exceeds an arbitrary threshold $S$ on a boundary, a new subdomain is created next to this boundary as shown on Figure 4. The value $S$ is generally fixed to 0 in this paper in order to detect any change of velocity on the boundaries of each subdomain.

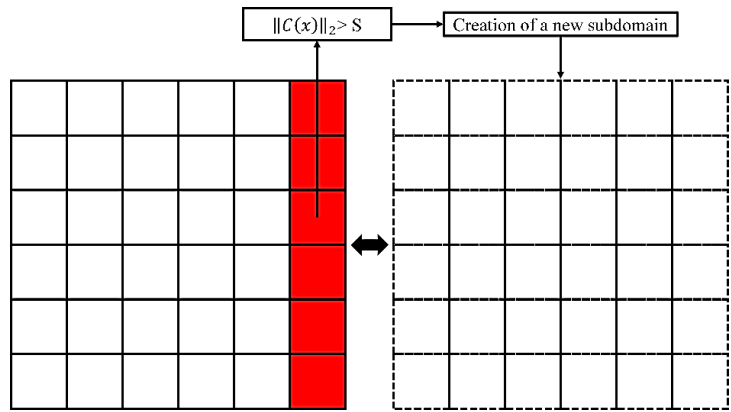

Figure 4: The criterion ‖$C_\alpha(x)$‖$_2$ is calculated on the boundary. If the criterion exceeds the threshold S then a new subdomain is created next to the boundary.

### 4.3. Algorithm

This section describes the algorithm for the multiphase and multicomponent lattice Boltzmann model with the inclusion of our progressive mesh algorithm. It is also useful in order to summarize the previous sections. The calculation of the criterion and the creation of new subdomains are achieved at the last step of the algorithm in order to not disturb the simulation process. Figure 5 describes our resulting algorithm.

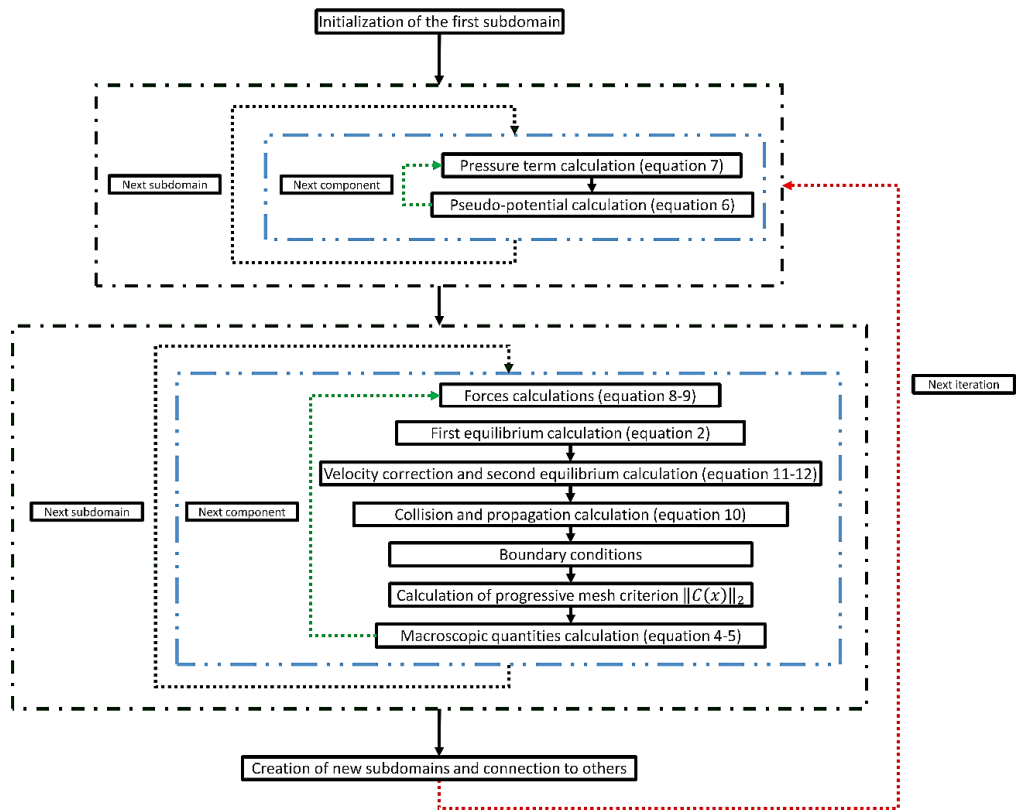

Figure 5: Algorithm for the multiphase and multicomponent Lattice Boltzmann model with the inclusion of our progressive mesh method. For colors, please refer to the PDF version of this paper.

## 4.4. Integration on Single-Node Multi-GPU Architecture

Efficiency of inter-GPU communications is surely the most difficult task in order to obtain good performance. Indeed, our simulations are composed of numerous subdomains which are added dynamically. The repartition of GPUs to the different subdomains is an important factor of optimization. An efficient assignment can have an important impact on the performance of the simulation. Indeed, it can reduce the communication time between subdomains and so reduce the simulation time.

### 4.4.1. Overlap Communications with Computations

Several data exchanges are needed for this type of model. The computation of interaction $F_{int}$ and inter-component $F_{ext}$ implies to have access to neighboring values of the pseudo-potential. The propagation step of LBM also implies to communicate several distribution functions $f_i$ between GPUs (Figure 6). Aligned buffers may be used for data transactions.

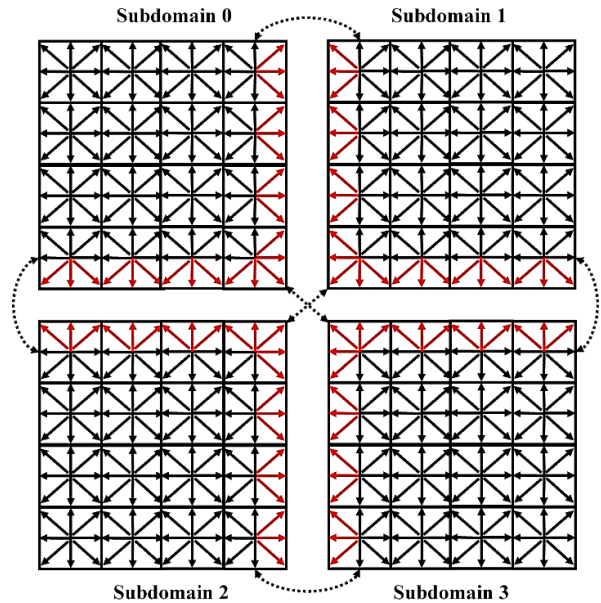

Figure 6: Schematic example for communication of distribution functions in 2D: red arrows corresponds to $f_i$ values to communicate between subdomains. For colors, please refer to the PDF version of this paper.

In order to obtain a simulation time as short as possible, it is necessary to overlap data transfer with algorithm calculations. Indeed, overlapping computations and communications allows to obtain a significant performance gain by reducing the waiting time of data. The idea is to separate the computation process into 2 steps: boundary calculations and interior calculations. Computations on the needed boundaries are firstly done. Communications between neighboring subdomains are also done while computing the interior. The different communications are thus performed simultaneously with calculations which allow good efficiency.

In most cases for lattice Boltzmann method, memory is transferred via zero-copy transactions to page-locked memory which allow good overlapping between communications and computations [17][13] [15].A different approach is studied in this paper concerning inter-GPU communications. In most recent HPC architectures, several GPUs can be connected to the same PCIe. To improve performance, Nvidia launched GPUDirect with CUDA 4.0.This technology allows to perform

Peer-to-Peer transfers and memory accesses between two compatible GPUs. The idea is to perform data transfer using Peer-to-Peer data transactions for GPUs sharing the same I/O hub and zero-copy transactions for others. This method allows to communicate data by bypassing the use of the CPU and therefore to accelerate the transfer (Figure 7). The use of this type of transaction improves performance and the efficiency of the simulation code.

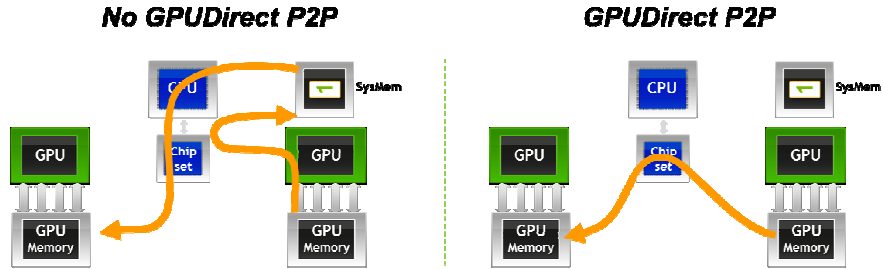

Figure 7: GPUDirect technology (source Nvidia).

### 4.4.2 Optimization of Data Transfer between GPUs

The repartition of GPUs is an important factor of optimization for this type of applications. Communications cost is generally a bottleneck for multi-GPU simulations. Three ways of data exchanges between sub domains are defined. A first assumption assumes that one sub domain is associated with one GPU. The first way concerns communications between sub domains belonging to the same GPU. In this case, the communication cost is extremely low because communications are performed on the same GPU global memory. The second and the third ways concern communications between sub domains belonging to different GPUs. A distinction is however made between Peer-to-Peer exchanges and zero-copy exchanges. This section has for goal to optimize dynamically the repartition of GPUs to new sub domains.

For a new sub domain $G$, the function $F$ is defined as follows:

$$F(G) = \sum_{G'} \gamma(G, G') \tag{14}$$

Where $G'$ denotes neighboring subdomains to $G$ and $\gamma(G, G')$ is defined as follows:

$$\gamma(G, G') = \begin{cases} 0 \; if \; GPU(G) = GPU(G') \\ 0.5 * sizeof(transfer) \; if \; GPU(G) \neq GPU(G') and \; GPU(G) \; can \; P2P \; GPU(G') \\ sizeof(transfer) \; if \; GPU(G) \neq GPU(G') and \; GPU(G) \; not \; P2P \; GPU(G') \end{cases} \tag{15}$$

The function $\gamma(G, G')$ compares the different ways of communications between the new subdomain and its neighbors. An arbitrary weighting value is included in order to promote Peer-to-Peer communications. The function $F$ performs the calculation of $\gamma$ for all active neighbors. The function $F(G)$ needs therefore to be minimized in order to obtain the best communication cost. This function is calculated for all available GPUs and the GPU with the minimum value is assigned to this subdomain. In order to keep load balancing, all GPUs have to be assigned dynamically and the same GPU could not be assigned two times as long as others GPUs are not assigned. Figure 8 explains via a simple example the principle of this optimization.

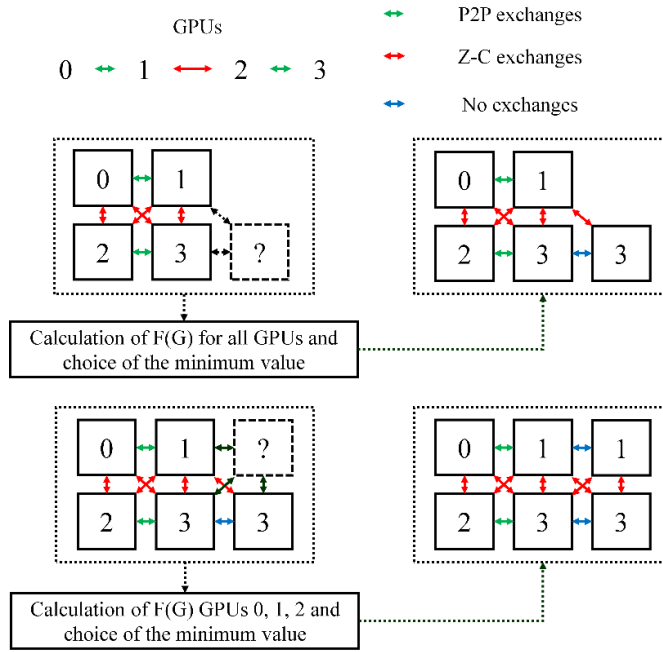

Figure 8: Schematic example in 2D for the optimization of the repartition of GPUs. The function $F(G)$ is calculated for all available GPUs and the GPU which have the minimum value is chosen. For colors, please refer to the PDF version of this paper.

### 5.1. Hardware

8 NVIDIA Tesla C2050 graphics cards Fermi architecture based machine are used to perform simulations. Table 1 describes some Tesla C2050 hardware specifications. Peer-to-Peer communications for our architecture are also described in Figure 9.

Table 1: Tesla C2050 Hardware specifications

| CUDA compute capability | 2.0 |
|---|---|
| Total amount of global memory | 2687 MBytes |
| (14) Multiprocessors, (32) scalar processors/MP | 448 CUDA cores |
| GPU clock rate | 1147 MHz |
| L2 cache size | 786432 bytes |
| Total amount of shared memory per block | 49152 bytes |
| Total number of registers available per block | 32768 |

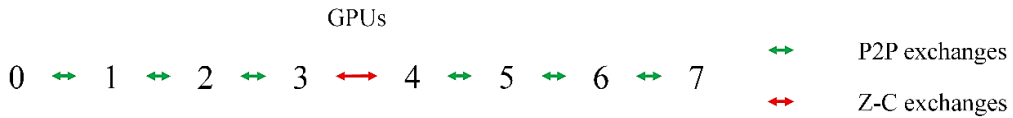

Figure 9: Peer-to-Peer communications accessibility for our architecture.

## 5.2. Simulations

Two simulations are considered on large simulation domain in order to evaluate the performance of our contribution. Both simulations include the use of two physical components. The geometry however differs between these simulations. The first simulation is based on a simple geometry composed of 1024*256*256 calculation cells where a fluid fills all simulation domains during the simulation (Figure 10). The second simulation is based on a complex geometry composed of 1024*1024*128 calculations cells where the fluid moves within channels (Figure 11).

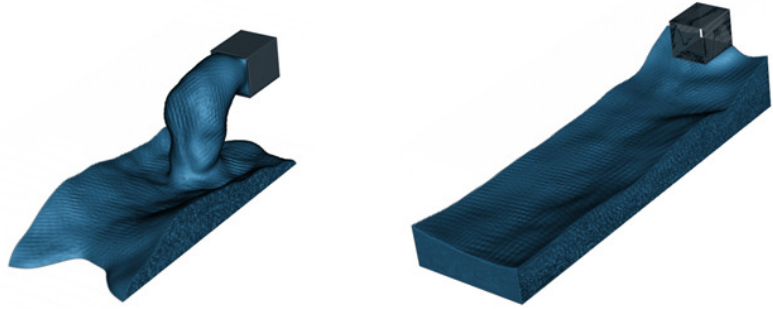

Figure 10: A two-component leakage simulation on a simple geometry with a domain size of 1024*256*256 cells.

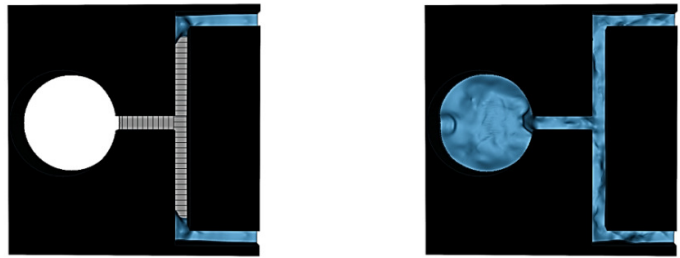

Figure 11: A two-component leakage simulation on a complex geometry composed of channels with a domain size of 1024*1024*128 cells.

## 5.3. Performance

This section deals with the performance obtained by our method. A comparison between the progressive mesh algorithm and the static mesh method generally used in literature is shown. The optimization of the repartition of GPUs on subdomains is also studied. The performance metric generally used for lattice Boltzmann method is the Million Lattice nodes Updates Per Second (MLUPS). It is calculated as follows:

$$Perf_{MLUPS} = \frac{domain\ size * number\ of\ iterations}{simulation\ time} \qquad (16)$$

This classical approach generally used in literature in order to perform simulations consists in equally dividing the simulation domain according to the number of GPUs. It offers generally good performance as communications can be overlapped with calculations. The use of Peer-to-Peer communications also has a beneficial effect on the performance, as shown on Figure 13. Peer-to-Peer communications allow obtaining a performance gain between 8 and 12% according to the number of GPUs used for the simulation described in Figure 10. Zero-copy communications offer a good scaling but an almost perfect scaling is obtained with the inclusion of Peer-to-Peer communications, as shown on Figure 12.

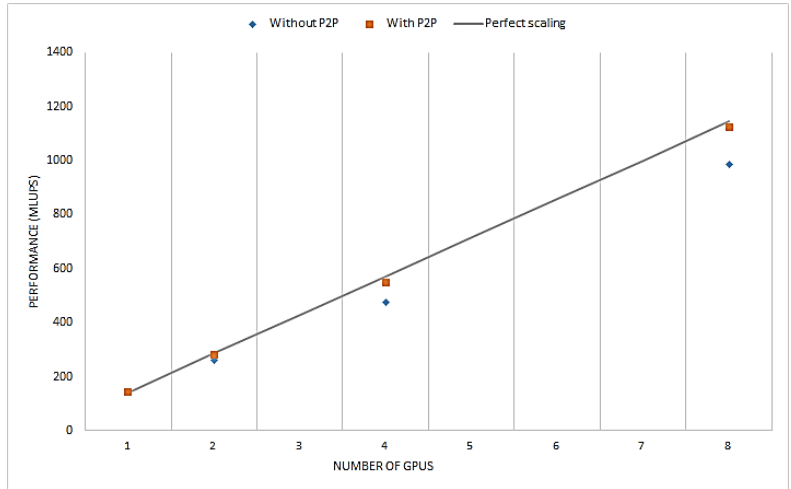

Figure 12: Comparison of performance between Peer-to-Peer communications with zero-copy communications for the simulation shown on Figure 10.

The inclusion of the progressive mesh also has an important beneficial effect on the simulation performance. Sub domains of size 128*128*128 are considered for these simulations. Figures 13 and 14 describes performance in terms of calculations and memory consumption for the simulation presented on Figure 10. Note that the progressive mesh algorithm obtains excellent performance at the beginning of the simulation. The addition of sub domains during the simulation has for consequence a decrease of performance until the convergence of the simulation. In this particular case, all simulation domain is meshed at the end of the simulation, as shown on Figure 14, which leads to a very slight decrease of performance compared to the static mesh. In terms of memory consumption, fast apparitions of news sub domains are noted which lead to have the entire simulation domain in memory after a few iterations.

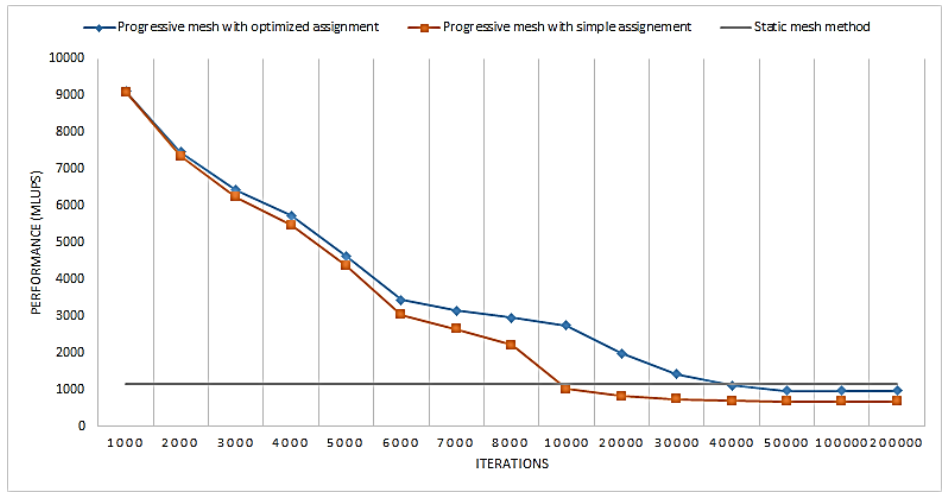

Figure 13: Comparison of performance between the progressive mesh method and the static mesh method for the simulation shown on Figure 10. The inclusion of the optimization for GPU assignment is also presented.

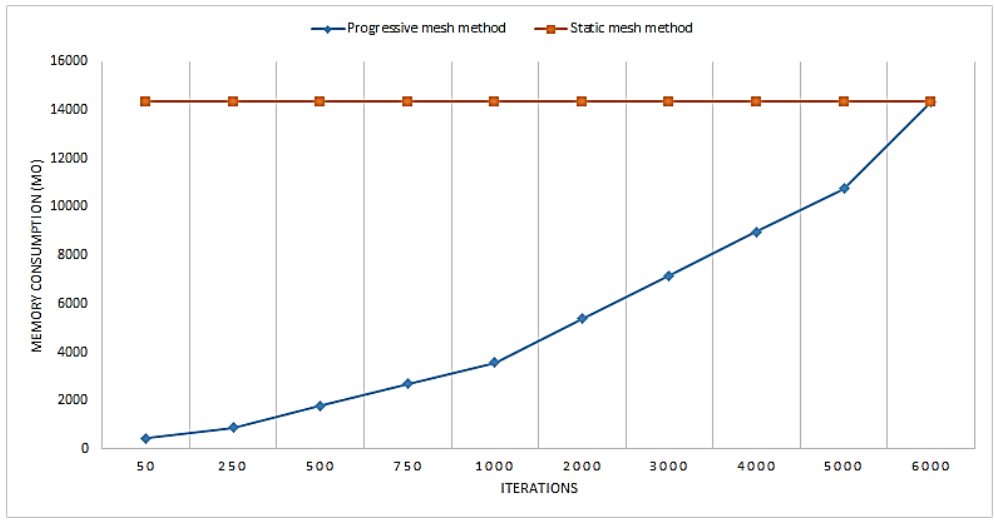

Figure 14: Comparison of memory consumption between the progressive mesh method and the static mesh method for the simulation shown on Figure 10.

Figure 13 also compares performance between two different assignments for GPUs. The first one is a simple assignation which assigns to new subdomain the first available GPU. The second one uses the optimization method presented in section 4.4.2. The comparison of these two methods leads to an important difference of performance. Indeed, a difference of approximatively 30% is noted at the convergence of this simulation between the two approaches. This difference is mostly due to the fact that the communication cost is more important for a simple assignment than an optimized assignment. Since subdomains are added dynamically and connected to each other, it is therefore important to optimize these communications in order to reduce the simulation time.

The same comparison is also done for the simulation presented on Figure 11, as shown on Figures 15 and 16. The main difference in this situation is the geometry of the simulation which is more complex and channelized. Physical simulations on channelized geometry are especially present on industrial structures.

In this case, the progressive mesh method shows excellent results. In terms of memory, this method is easily able to simulate on a global simulation domain of size 1024*1024*128 and more while the static mesh method is unable to perform the simulation. The amount of needed memory is indeed too important for this simulation. Figure 15 shows the evolution of memory consumption during the simulation. The memory cost at the convergence of the simulation is far less important than the static mesh method. A gain of approximatively 50% of memory is noted for this particular simulation. This is due to the fact that the progressive mesh method automatically adapts to the evolution of the simulation and so only needed zones of the global simulation domain are meshed.

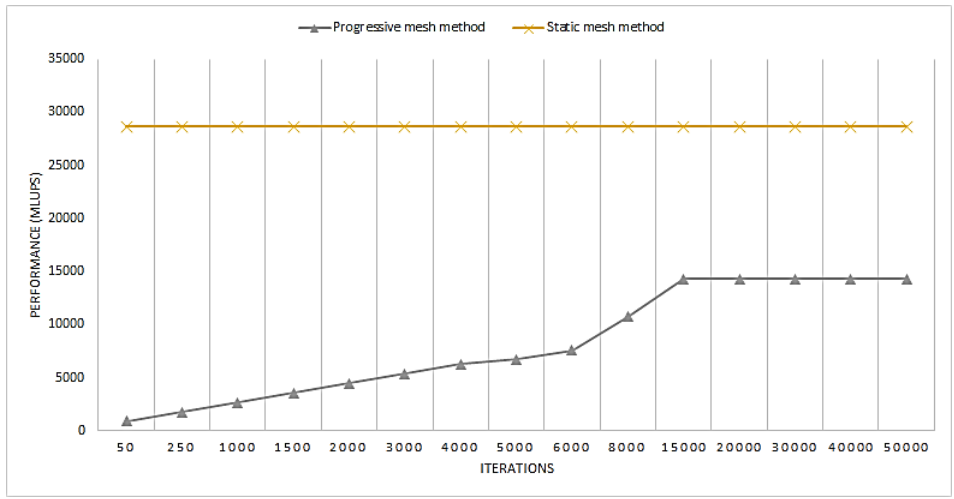

Figure 15: Comparison of memory consumption between the progressive mesh method and the static mesh method for the simulation shown on Figure 11.

The comparison of the repartition of GPUs is also described in Figure 16. An important performance gain (19%) is still noted for this simulation. This proves that a dynamic optimization method is important in order to obtain good performance. Moreover, the fact that the domain does not need to be fully meshed brings an important gain in performance. The geometry has therefore an important impact on the performance on the progressive mesh method.

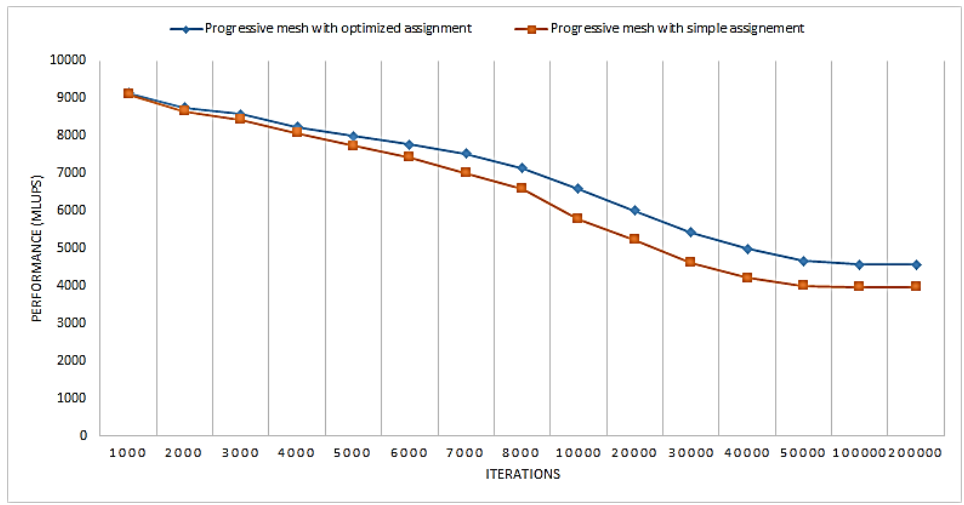

Figure 16: Comparison of performance between a simple repartition of GPUs with an optimized assignment of GPUs for the simulation shown on Figure 11.

## 6. CONCLUSION

In this paper, an efficient progressive mesh algorithm for physical simulations using the lattice Boltzmann method is presented. This progressive mesh method can be a useful tool in order to perform several types of physical simulations. Its main advantage is that subdomains are automatically added to the simulation by the use of an adapted criterion. This method is also able to save a lot of memory and calculations in order to perform simulations on large installations.

The integration of the progressive mesh method on single-node multi-GPU architecture is also treated. A dynamic optimization of the repartition of GPUs to subdomains is an important factor in order to obtain good performance. The combination of all these contributions allows therefore performing fast physical simulations on all types of geometry. The progressive mesh method is therefore an interesting alternative because it allows obtaining similar or better performances than the usual static mesh method.

The progressive mesh algorithm is however limited to the memory of the GPU which is generally far more inferior to the CPU RAM. The creation of new subdomains is indeed possible while there is a sufficient amount of memory on the GPUs. Extensions of this work to cases that require more memory than all GPUs can handle is now under investigation. Data transfer optimizations with the CPU host will therefore be essential to keep good performances.

## ACKNOWLEDGEMENTS

This work has been made possible thanks to collaboration between academic and industrial groups, gathered by the INNOCOLD association.

**AUTHORS**

Julien Duchateau is a PhD student in computer science at the Université du Littoral Côte d'Opale in France. His main research interest are massive parallelism on CPUs and GPUs, physical simulations and computer graphics.

François Rousselle is an associate professor in computer science at the Université du Littoral Côte d'Opale in France. His main research interests are computer graphics, physical simulations, virtual reality and massive parallelism.

Nicolas Maquignon is a PhD student in simulation and numerical physics at the Université du Littoral Côte d'Opale. His main research interests are numerical physics, numerical mathematics and numerical modeling.

Christophe Renaud is a professor in computer science at the Université du Littoral Côte d'Opale in France. His main research interests are computer graphics, virtual reality, physical simulations and massive parallelism.

Gilles Roussel is an associate professor in automatic at the Université du Littoral Côte d'Opale in France. His main research interests are automatic, signal processing, physical simulations and industrial computing.